# TSPTQ-ViT: TWO-SCALED POST-TRAINING QUANTIZATION FOR VISION TRANSFORMER

*Yu-Shan Tai, Ming-Guang Lin, and An-Yeu (Andy) Wu*

Graduate Institute of Electrical Engineering, National Taiwan University, Taipei, Taiwan

{clover, chrislin}@access.ee.ntu.edu.tw, andywu@ntu.edu.tw

## ABSTRACT

Vision transformers (ViTs) have achieved remarkable performance in various computer vision tasks. However, intensive memory and computation requirements impede ViTs from running on resource-constrained edge devices. Due to the non-normally distributed values after Softmax and GeLU, post-training quantization on ViTs results in severe accuracy degradation. Moreover, conventional methods fail to address the high channel-wise variance in LayerNorm. To reduce the quantization loss and improve classification accuracy, we propose a two-scaled post-training quantization scheme for vision transformer (TSPTQ-ViT). We design the value-aware two-scaled scaling factors (V-2SF) specialized for post-Softmax and post-GeLU values, which leverage the bit sparsity in non-normal distribution to save bit-widths. In addition, the outlier-aware two-scaled scaling factors (O-2SF) are introduced to LayerNorm, alleviating the dominant impacts from outlier values. Our experimental results show that the proposed methods reach near-lossless accuracy drops (<0.5%) on the ImageNet classification task under 8-bit fully quantized ViTs.

## 1. INTRODUCTION

Originating from natural language processing (NLP) tasks [1], transformer-based models have received fabulous performance and outperformed convolutional neural networks (CNNs) in various computer vision (CV) tasks [2]-[6]. However, vision transformers (ViT) suffer from heavier memory and computational costs than CNNs. For example, there are 307 M parameters and 64 G FLOPs in ViT-L [2]. The unaffordable overheads hinder ViTs from running on resource-constrained edge devices, confining their real-world applications. Consequently, model compression for ViTs arises as an urgent problem needed to be solved immediately.

There are many model compression techniques, such as quantization [7]-[16], pruning [17][18], and dimension reduction [19]-[22]. Without modifying the model architecture, quantization saves bit-widths by mapping floating-point (FP) values to discrete integers. Existing quantization works comprise two fields, quantization-aware training (QAT) and post-training quantization (PTQ). QAT [11]-[13] shows robustness under aggressive quantization by adding quantization loss during training. However, these methods suffer from intolerable fine-tuning costs. For instance, TerViT [13] reaches 2-bit quantization at the cost of fine-tuning for 300 epochs. To compress models without extra fine-tuning, PTQ [14]-[16] leverages a small amount of unlabeled data to calibrate scaling factors, significantly improving the efficiency.

Although PTQ for CNNs has been exhaustively researched previously [7]-[10], directly transferring these techniques to

This research work is financially supported in part by Novatek, under grant 110HT945009, and in part by Ministry of Science and Technology, Taiwan, under grants MOST 111-2218-E-002-018 -MBK and MOST 110-2221-E-002 -184 -MY3.

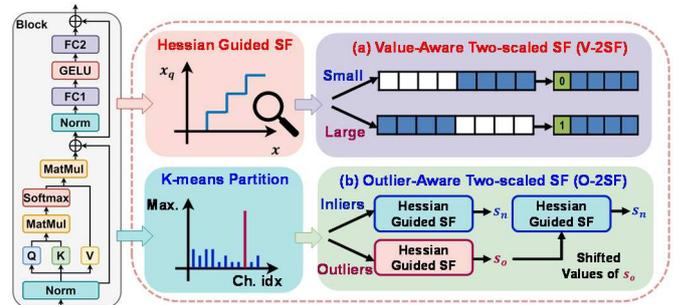

**Fig. 1.** Overview of the proposed TSPTQ-ViT: (a) V-2SF and (b) O-2SF, where SF denotes scaling factors.

ViTs ends up with dramatic accuracy degradation [14]. The authors of PTQ4ViT [15] point out that the accuracy drop comes from the non-normal distribution after Softmax and GeLU. Thus, they [15] design a region-specific quantization scheme to reduce the quantization loss. On the other hand, most works compute Softmax and LayerNorm with FP values to avoid catastrophic accuracy drops, failing to achieve fully quantized transformers [14][15]. FQ-ViT [16] is the first framework to achieve PTQ for fully quantized ViT. However, the channel-wise scaling factors design introduces extra memory overhead.

The reasons that make PTQ challenging to deploy on ViTs can be summarized as follows:
1) *Non-normal distribution after Softmax and GeLU:* The values after Softmax present an unbalance distribution among [0,1]; most values are close to 0 while a few are near 1. On the other hand, post-GeLU values exhibit an asymmetrical distribution, where the positive range is far broader than the negative one. Directly applying uniform quantization on these non-normally distributed values leads to unrecoverable information loss.
2) *High channel-wise variance in LayerNorm*: There are high inter-channel variances among the inputs of LayerNorm, where the maximum is 40× larger than the median. The outlier values dominate the magnitude of scaling factors, thus severely sacrificing the precision of small values.

To address the above issues, we propose a two-scaled post-training quantization scheme for vision transformers (TSPTQ-ViT), as shown in Fig. 1. Our main contributions are as follows:
1) *Value-Aware Two-Scaled Scaling Factors (V-2SF)*: We propose a two-scaled strategy specialized for post-Softmax and post-GeLU values, exploiting bit sparsity in non-normal distribution.
2) *Outlier-Aware Two-Scaled Scaling Factors (O-2SF)*: We design another two-scaled mechanism operated on the input channel dimension of LayerNorm, alleviating the dominant impact from outlier values.
3) With extensive experiments on different vision transformers, our proposed TSPTQ-ViT achieves near-lossless performance under 8-bit fully quantized ViTs.

The rest of this paper is organized as follows. Section **2** briefly introduces vision transformers and existing quantization methods. Section **3** illustrates the proposed TSPTQ-ViT, which includes V-2SF and O-2SF. The experiments and analyses are illustrated in Section **4**. Finally, Section **5** concludes our work.

## 2. RELATED WORK

### 2.1. Vision Transformer

Transformer-based models are first applied in NLP tasks [1] and then successfully transferred to the CV field, such as image classification [2]-[4], object detection [5], and semantic segmentation [6]. Unlike CNNs, which extract local information from neighboring pixels, vision transformers learn global information through self-attention modules [23] and achieve superior performance. ViT [2] is the first work that directly applies a transformer on image patches, obtaining impressive performance on multiple image classification tasks. However, to attain powerful results, ViTs require a long training time and substantial computational cost, which is quadratic with the sequence length of image patches. Consequently, the ViTs proposed recently aim to improve efficiency. The authors of DeiT [3] introduce a teacher-student strategy specific to transformers, achieving competitive accuracy with less training data. As for Swin [4], the authors apply a hierarchical architecture and compute self-attention within the shifted local windows, reducing computational complexity and enhancing accuracy. Unlike the above works, which modify training strategy or network architectures, this paper focuses on model quantization. The proposed methods can be applied to different transformer-based models.

### 2.2. Model Quantization

Quantization is prevalently used for different compression tasks. The uniform symmetric quantization is the most common strategy, which projects floating-point values $x$ to $b$-bit integers $x_{q,b}$ by a uniform scaling factor $s$ and a zero-point $zp$:

$$x_{q,b} = clamp(\left\lfloor \frac{x}{s} \right\rfloor, -2^{b-1}, 2^{b-1} - 1), \qquad (1)$$

$$s = \frac{|\max(x) - zp|}{2^b - 1}, zp = 0, \qquad (2)$$

where $clamp(x,l,h)$ is used to clamp x in $[l,h]$. Existing quantization methods can be divided into two fields: quantization-aware training (QAT) and post-training quantization (PTQ). In QAT [11]-[13], a quantization-aware loss is introduced during the optimization of model parameters, thus enhancing model robustness under aggressive quantization. The authors of Q-ViT [12] utilize the differentiable strategy to maintain model accuracy under 3-bit quantization. Moreover, the authors of TerViT [13] use progressive quantization with fine-tuning for 300 epochs to reach 2-bit quantization.

Though QAT achieves low bit-width design, it often requires an entire labeled dataset, long training periods, and hyper-parameter tuning. On the contrary, PTQ [14]-[16] only leverages a small amount of unlabeled data to determine the optimal scaling factors without additional fine-tuning. The authors of [14] first introduce PTQ to ViTs, combining a ranking loss and a similarity metric to determine the suitable scaling factors. However, to avoid significant accuracy degradation, this work [14] does not quantize Softmax and LayerNorm.

The small values after Softmax and GeLU are suppressed to zero under uniform quantization, thus leading to huge quantization error. Consequently, the authors of PTQ4ViT [15] design a region-specific quantization method. For post-Softmax values, the two regions are $R1 = [0, 2^{b-1}s_{R1})$ and $R2 = [2^{b-1}s_{R1}, 1]$, and the corresponding scaling factors are set as $s_{R1} = 2^{-m}s_{R2}$ and $s_{R2} = 1/2^{b-1}$. To determine the optimal values of $m$, the authors use a Hessian guided metric to search with the scaling factor of weights alternatively. As for post-GeLU values, $s_{R1}$ and $s_{R2}$ are used to quantize negative and positive values, respectively. They set $s_{R1} = 2^{-m}s_{R2}$ and the search space of $s_{R2}$ comprises $N$ linearly divided values in $[0, 1.2 * \frac{max}{2^{b-1}}]$. Then, they apply the Hessian guided metric to search the two scaling factors simultaneously. Although [15] successfully reduces quantization loss of post-Softmax and post-GeLU values, some issues still remain to be solved. First, the fixed design of $s_{R2}=1/2^{b-1}$ for post-Softmax leads to redundant integer bins if the maximum value is far less than 1. Secondly, though negative small post-GeLU values are well-preserved, most positive small values are still mapped to zero. Lastly, PTQ4ViT [15] computes Softmax and LayerNorm with FP values and fails to achieve fully quantized ViTs.

FQ-ViT [16] is the first framework to achieve PTQ for fully quantized ViTs. The authors utilize a polynomial approximation and log2 quantization for Softmax. Moreover, they apply the channel-wise scaling factors to solve the high channel-wise variance in LayerNorm. In each layer, there are four candidates $s_i$ for channel-wise selection, which can be specified as:

$$C_s = \{s_i = s_{i-1} \gg 1 | i = 1,2,3, s_0 = s\}. \qquad (3)$$

Then, the authors select the final scaling factors with minimum L2 loss between quantized and unquantized values. Although achieving fully quantized ViTs, there is still room for improvement. First, log2 quantization enhances the precision of small values but introduces severe quantization error for large values. Next, the channel-wise scaling factors require additional 2 bits to indicate the index $i$ of selected $s_i$ for each channel, introducing additional memory costs. Lastly, the candidates $C_s$ are highly affected by the outlier values, leading to a sub-optimal result.

To alleviate the above issues of prior works, we propose a novel PTQ framework for ViTs, TSPTQ-ViT, achieving near-lossless accuracy with neglectable memory overheads.

## 3. PROPOSED TWO-SCALED POST-TRAINING QUANTIZATION FOR VISION TRANSFORMER

### 3.1. Value-Aware Two-Scaled Scaling Factors (V-2SF)

To reduce quantization loss of post-Softmax and post-GeLU values, we propose the value-aware two-scaled scaling factors (V-2SF) to utilize bit sparsity of the non-normal distribution. Inspired by [24], leveraging the most significant consecutive bits, to store the activations in CNNs, we further extend this idea to ViTs, as shown in Fig. 2. For large values, we only store the most significant few bits. As for small values, we reserve the least significant few bits instead. Therefore, the scaling factors $s_l$ for large values can be fast aligned by shifting scaling factors $s_s$ of small values, e.g., $s_l = 2^{-m} s_s$. We empirically find out $m = 4$ and $m = 3$ are the optimal configurations for values after Softmax and GeLU, respectively.

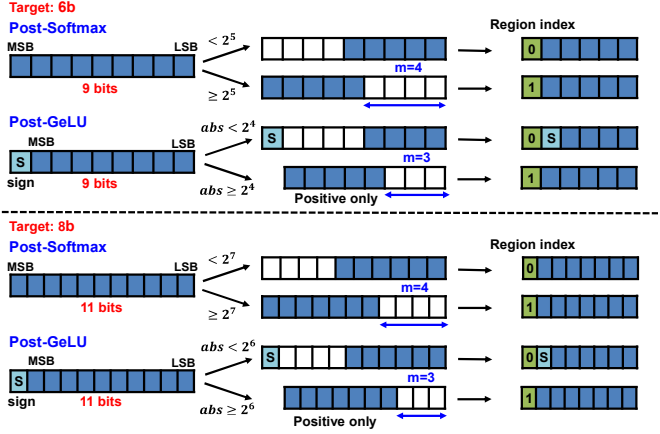

**Fig. 2.** Value-aware two-scaled scaling factors (V-2SF) under (a) 6-bit and (b) 8-bit quantization.

We first elaborate on the V-2SF for post-Softmax values, which only contain unsigned values. Take 6-bit quantization for example. The final format includes 1 bit for the region index and 5 bits for the quantized values. Therefore, the initial quantization bit-width is set as 9-bit, and we use the Hessian guided metric [15] to decide the scaling factor $s_s$. Afterwards, we separate the values into two regions according to their quantized values. If the values are less than $2^5$, five least significant bits are sufficient to represent the original values. As for those values equal to or larger than $2^5$, we record the five most significant bits while rounding the first truncated bit. Finally, with 1 bit for the region index to indicate using $s_l$ or $s_s$, the final bit-width is 6 bits.

As for the V-2SF for post-GeLU values, there is a little difference due to the existence of sign bit. In the 6-bit quantization case, we initialize the values to 9-bit quantization, where the first bit is the sign bit and the others denote the magnitude. Then, we evaluate the absolute quantized values to divide them into two regions. If the absolute values are less than $2^4$, the least significant 4 bits are enough to reconstruct the original values. Besides the region index, a sign bit is also required in this case. Thus, the total bit-width is 6 bits. As for the other case, absolute values are equal to or larger than $2^4$. Note that in post-GeLU values, the range of positive values is far broader than that of negative ones, the values with large magnitudes are all positive. Consequently, we can remove the sign bit and use the five most significant bits plus one region index to denote the values, which is 6 bits in total.

The flow of 8-bit quantization is similar to that of 6-bit quantization, and our V-2SF can extend to any bit-width by following the above rules. By exploiting the bit sparsity in non-normal distribution, the proposed V-2SF effectively reduces the quantization loss of post-Softmax and post-GeLU values. In Sec. **4.2**, we would further visualize the post-quantization distribution to compare our V-2SF with PTQ4ViT [15].

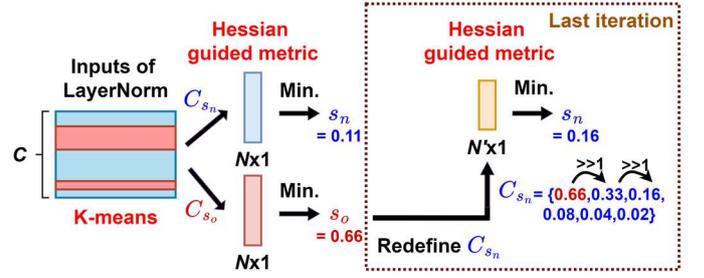

**Fig. 3.** Outlier-aware two-scaled scaling factors (O-2SF).

### 3.2. Outlier-Aware Two-Scaled Scaling Factors (O-2SF)

In this section, we propose the outlier-aware two-scaled scaling factors (O-2SF) to alleviate the high channel-wise variance issue. The flowchart is shown in Fig. 3. We observe that extreme values only appeared in a few channels. Consequently, rather than resort to channel-wise scaling factors search [16], we identify the outlier channels and assign them a scaling factor $s_o$ different from the scaling factor $s_n$ of others.

To detect the outliers, we operate the K-means algorithm on the absolute maximum of total $C$ channels layer-wisely. Then, we independently find the Hessian guided scaling factors [15] from $N$ linearly divided candidates $C_{s_o}$ and $C_{s_n}$ for $s_o$ and $s_n$, respectively. As mentioned in Sec. **2**, the Hessian guided scaling factors are alternatively searched to co-optimize with the scaling factors of weight. For fast alignment, $s_o$ should be obtained by shifting $s_n$. Therefore, in the last searching iteration, we redefine the candidates $C_{s_n}$ for $s_n$ as follows:

$$C_{s_n} = \{s_o \gg k | k = [0, N']\}. \tag{4}$$

After that, as the same as previous iterations, we calculate the Hessian guided metric for the $N'$ candidates and select the one with minimum as the final $s_n$.

Compared with FQ-ViT [16], which requires storing extra 2 bits per channel, we just need 1 bit per channel to specify using $s_o$ or $s_n$. Moreover, by introducing the Hessian guided searching, we can avoid the dominance of the outlier values and obtain the optimal scaling factors. In Sec. **4**, we would further compare the performance of our method with prior works to validate its effectiveness.

### 4. SIMULATION RESULTS

In the following experiments, we implement our proposed method on ImageNet (ILSVRC 2012) [25] with pre-trained ViT [2], DeiT [3], and Swin [4] by timm [26]. We randomly sample 32 images from training data for calibration. The number of Hessian guided search rounds is 3, and the size of candidates $N/N'$ is 100/6. Without specifying, the search space of candidates is set as $[0, 1.2 * \frac{max.}{2^{b-1}}]$ as in [15]. We quantize all weights and inputs for the LayerNorm and fully connected layers including the first and the last layers. We also quantize Softmax, and utilize integer-only Softmax [27] to approximate the exponential function. We apply Hessian guided scaling factors [15] to quantize all weights and activations. Our TSPTQ-ViT finally achieves integer-only inference and fully quantized ViT models.

**Table 1.** Comparison of the top-1 accuracy with state-of-the-art methods on ImageNet dataset. * indicates mixed-precision is applied, *W/A* denotes the bit-width for weight/activation, *# img* stands for the size of calibration data, *Fully Quant.* indicates whether the method supports fully quantization.

| Method | W/A | # img | Fully Quant. | ViT-S | ViT-B | ViT-L | DeiT-T | DeiT-S | DeiT-B | Swin-T | Swin-S | Swin-B |
|---|---|---|---|---|---|---|---|---|---|---|---|---|
| FP | 32/32 | ✗ | ✗ | 81.39 | 84.53 | 85.84 | 72.18 | 79.85 | 81.99 | 81.37 | 83.21 | 85.27 |
| Liu. [14] | 8/8 | 1024 | ✗ | - | 76.98* | 76.41 | - | 77.47 | 80.48 | - | - | - |
| PTQ4ViT [15] | 8/8 | 32 | ✗ | 81.00 | 84.25 | 85.75 | 71.63 | 79.47 | 81.48 | 81.25 | 83.11 | 85.15 |
| **V-2SF** | 8/8 | 32 | ✗ | **81.27** | **84.25** | **85.82** | **71.98** | **79.74** | **81.77** | **81.26** | **83.11** | **85.19** |
| Liu. [14] | 6/6 | 1024 | ✗ | - | 75.26* | 75.46* | - | 74.58 | 77.02 | - | - | - |
| PTQ4ViT [15] | 6/6 | 32 | ✗ | 78.63 | 81.65 | 84.79 | 69.62 | 76.28 | 80.25 | 80.47 | 82.38 | 84.01 |
| **V-2SF** | 6/6 | 32 | ✗ | **79.34** | **82.01** | **85.20** | **70.68** | **77.27** | **80.25** | **80.68** | **82.41** | **84.12** |
| FQ-ViT [16] | 8/8 | 1000 | ✓ | - | 83.31 | 85.03 | 71.61 | 79.17 | 81.20 | 80.51 | 82.71 | 82.97 |
| V-2SF | 8/8 | 32 | ✓ | 70.74 | 83.53 | 85.53 | 71.83 | 79.48 | 80.80 | 80.95 | 82.61 | 80.99 |
| **V-2SF+O-2SF** | 8/8 | 32 | ✓ | **81.20** | **84.11** | **85.81** | **71.87** | **79.56** | **81.72** | **81.19** | **83.07** | **85.11** |

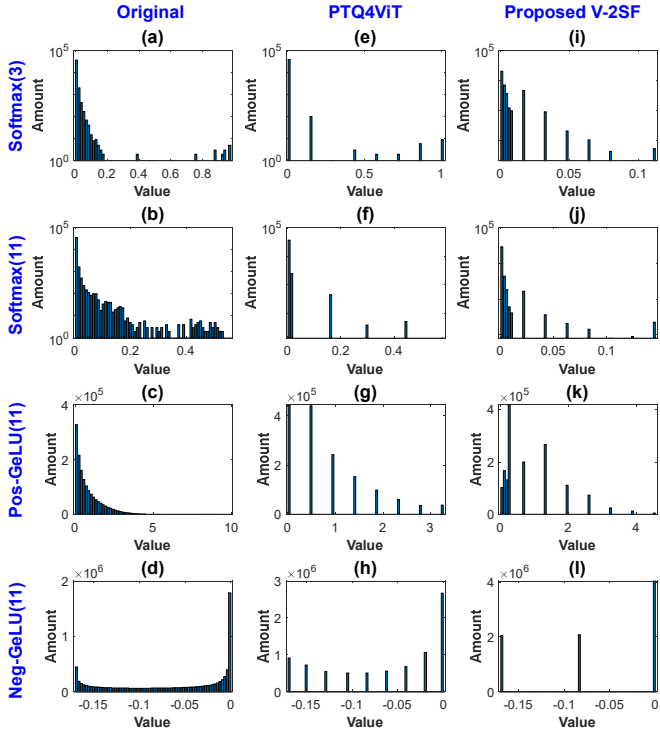

**Fig. 4.** Distribution of post-Softmax values, positive post-GeLU values and negative post-GeLU values of ViT-S under 4-bit quantization: (a)~(d) original, (e)~(h) PTQ4ViT [15], (i)~(l) proposed V-2SF. The numbers in parentheses denote the block index.

### 4.1. Comparison with State-of-The-Art PTQ Methods

In this section, we compare the top-1 accuracy among different methods, as shown in Table. 1. Compared with methods without fully quantization, the proposed V-2SF reaches similar or even higher accuracy than PTQ4ViT [15]. In 6-bit quantization of DeiT-T and DeiT-S, our V-2SF outperforms PTQ4ViT with approximately 1% accuracy. To further analyze their difference, the post-quantization distribution of V-2SF and PTQ4ViT will be demonstrated in the next section. As for the fully quantization scenario, the proposed V-2SF+O-2SF obtains higher accuracy than FQ-ViT [16] in all types of ViTs. For Swin-B, we even improve 2.14% accuracy than FQ-ViT. Also note that, as mentioned in Sec. **3**, the memory overhead of our O-2SF is 1 bit per channel, while FQ-ViT requires 2 bits per channel. Moreover, to validate the effectiveness of O-2SF, we also compare V-2SF+O-2SF with V-2SF. V-2SF directly applies Hessian guided scaling factors without K-means partition. The simulation results demonstrate that introducing O-2SF indeed enhances accuracy and proves special processing for outliers is necessary.

In summary, our TSPTQ-ViT achieves fully quantized ViTs with less than 0.5% accuracy drop under 8-bit weight/activation quantization, achieving higher accuracy than prior state-of-the-art PTQ methods.

### 4.2. Visualization of Post-Quantization Distribution

To compare our V-2SF with PTQ4ViT [15], we visualize their post-quantization distribution in Fig. 4. Observe the distribution of post-Softmax values under PTQ4ViT in Fig. 4(e) and (f), we can notice the distribution of Fig. 4(f) is sparser than that of Fig. 4(e). Since the maximum value of the original (Fig. 4(b)) is only 0.56, the fixed design of $s_{R2}=1/2^{b-1}$, which is able to present values up to 1, leads to redundant integer bins. However, since we apply Hessian guided scaling factors, the step size of V-2SF is flexible. Moreover, the maximum values shown in Fig. 4(i) and (j) are roughly 0.1, implying the flexible design in V-2SF can clip large values and enhance the granularity of the remaining ones. Fig. 4(g) and (h) show the distribution of post-GeLU values under PTQ4ViT. Though region-specific strategy keeps negative small values well-preserved as in Fig. 4(h), most positive small values are still mapped to zero as in Fig. 4(g). As for the proposed V-2SF, small positive values share the same scaling factors $s_s$ with negative values. Thus, rather than over-designed negative region, we improve the precision of small positive values, as shown in Fig.4 (k) and (l). In summary, considering the characteristic of the non-normal distribution, our V-2SF can better preserve model accuracy than prior works.

## 5. CONCLUSION

In this paper, we introduce a two-scaled PTQ design for ViTs, TSPTQ-ViT, to reduce the quantization loss of ViTs and improve accuracy. Specifically, we design the V-2SF specialized for post-Softmax and post-GeLU values, exploiting the bit sparsity in non-normally distributed values. Furthermore, we develop the O-2SF targeted for LayerNorm, identifying the outlier channels and assigning them a different scaling factor from others. Experimental results show our TSPTQ-ViT reaches near-lossless accuracy under 8-bit fully quantized ViTs.